\documentclass[aps,twocolumn,showpacs,pra,groupedaddress,superscriptaddress,nofootinbib]{revtex4}
\usepackage{graphicx,amsmath,amssymb,amsbsy,subfigure,hyperref,bbm,times,txfonts}
\begin{document}

\addtolength{\textheight}{0.5cm}
\newcommand{\ket}[1]{|#1\rangle}
\newcommand{\bra}[1]{\langle#1|}
\newcommand{\ketbra}[1]{| #1\rangle\!\langle #1 |}
\newcommand{\kebra}[2]{| #1\rangle\!\langle #2 |}
\newcommand{\id}{\mathbbm{1}}
\newcommand{\ohm}{\Omega_{\rm CQ}}
\newcommand{\rhobd}{\rho^{\vec{c}}_{AB}}
\providecommand{\tr}[1]{\text{tr}\left[#1\right]}
\providecommand{\tra}[1]{\text{tr}_A\left[#1\right]}
\providecommand{\trb}[1]{\text{tr}_B\left[#1\right]}
\providecommand{\abs}[1]{\left|#1\right|}
\providecommand{\sprod}[2]{\langle#1|#2\rangle}
\providecommand{\expect}[2]{\bra{#2} #1 \ket{#2}}

\title{Comparative investigation of the freezing phenomena
           for quantum correlations under nondissipative decoherence}
\author{Benjamin Aaronson}
\affiliation{School of Mathematical Sciences, The University of Nottingham, University Park, Nottingham NG7 2RD, United Kingdom}
\author{Rosario Lo Franco}
\affiliation{Dipartimento di Fisica e Chimica, Universit\`a di Palermo, via Archirafi 36, 90123 Palermo,
Italy}
\author{Gerardo Adesso}
\affiliation{School of Mathematical Sciences, The University of Nottingham, University Park, Nottingham NG7 2RD, United Kingdom}
\begin{abstract}
We show that the phenomenon of frozen discord, exhibited by specific classes of two-qubit states under local nondissipative decoherent evolutions, is a common feature of all known {\it bona fide} measures of general quantum correlations. All those measures, despite inducing typically inequivalent orderings on the set of nonclassically correlated states, return a constant value in the considered settings. Every communication protocol which relies on quantum correlations as resource will run with a performance completely unaffected by noise in the specified dynamical conditions. We provide a geometric interpretation of this phenomenon.
\end{abstract}
\date{May 31, 2013}
\maketitle

\section{Introduction}
Quantum correlations, seminally quantified by the quantum discord \cite{OZ,HV}, stand as one of the most general manifestations of nonclassicality in composite systems. They can be revealed in the process of locally measuring a subsystem, even in states where entanglement or nonlocality are absent. Despite a massive surge in recent studies investigating interpretation, quantification, and applications of discord and related quantifiers of quantum correlations \cite{modirev}, it is a fact that these quantities remain far less understood than entanglement \cite{entanglement}. Few properties are now set in stone as necessary requirements to identify a quantifier, say ${\cal Q}$, as a {\it bona fide} measure of general quantum correlations, revealed on the subsystem $A$, in bipartite states $\rho_{AB}$. We list them below: \par
{\it P1.}~Vanishing on classical-quantum (CQ) states: ${\cal Q}(\rho_{AB})\geq 0$ for all $\rho_{AB}$ and ${\cal Q}(\chi_{AB})=0$ for all states $\chi_{AB}$ belonging to the set $\ohm$ of CQ states,  $\chi_{AB} = \sum_i p_i \ketbra{i}_A \otimes {\tau_i}_B$, with $\{p_i\}$ being a probability distribution,  ${\ket{i}_A}$ an orthonormal basis for subsystem $A$, and ${\tau_i}_B$ arbitrary states for subsystem $B$; \par
{\it P2.}~Invariance under local unitaries: ${\cal Q}\big((U_A \otimes U_B) \rho_{AB}(U_A \otimes U_B)^{\dagger}  \big)= {\cal Q}(\rho_{AB})$; \par
{\it P3.}~Nonincreasing under  local operations on the unmeasured party $B$: ${\cal Q}\big((\id_A \otimes \Phi_B)[\rho_{AB}]\big) \leq {\cal Q}(\rho_{AB})$, where $\Phi_B$ is a completely positive and trace preserving map (i.e., a quantum channel) acting on subsystem $B$; \par
{\it P4.}~Reduction to entanglement on pure states: ${\cal Q}(\ketbra{\psi_{AB}})$ is an entanglement monotone \cite{entanglement}.

These criteria are certainly not complete, as in particular a definition of the operations on $A$, or of the allowed classical communication between $A$ and $B$, under which discord and any valid ${\cal Q}$ should not increase, has not been accomplished \cite{noncomm,behavior,meznaric,modirev}.  In the quest to unveil the most essential signatures of quantumness in complex systems, it is then of wide interest to identify physically insightful properties that underly the notion of  quantum correlations as opposed to entanglement, and should then be reflected by any valid measure thereof. One such property is, for instance, the absence of monogamy \cite{arequantum}. This area of investigation has also a strong technological motivation \cite{merali}: it is believed, and in some case proven, that states with quantum correlations other than entanglement can be employed as resources for several quantum computation \cite{datta,dqc1expwhite,dqc1explaflamme}, communication \cite{madhok:pr2011a,cavalcanti:pr2011a,np1,np2}, and metrology setups \cite{modix,lqu}. Identifying the  distinctive traits of such correlations, in particular with respect to their resilience under noise, might lead to valuable recipes for their practical exploitation.

Numerous works have in fact investigated the dynamics of general quantum correlations in open quantum systems undergoing various types of Markovian or non-Markovian evolutions, as recently reviewed in \cite{modirev,revbraz,revsaro}. One evident feature is that discord is typically more robust than entanglement and does not suffer from sudden death issues \cite{maziero,acinferraro}; discord can even be created by local operations on the measured party $A$  (nonunital channels if $A$ is a qubit) \cite{behavior,ciccagio}. A particularly fascinating phenomenon can occur for two-qubit states undergoing nondissipative decoherence: their discord can remain constant, or {\it frozen}, for an interval of time in Markovian conditions \cite{mazzola}. A forever frozen discord \cite{pigna}, or multiple intervals of recurring frozen discord \cite{lauraint,sarophys,sarorevival}, can further occur when the dynamics is non-Markovian. Necessary and sufficient conditions for the freezing have been derived in \cite{necsuff}. However, to date such a feature seemed bound to the choice of particular (mainly entropic) quantifiers of discord \cite{modirev}. It is natural to question whether freezing just happens as a mathematical accident, or whether it bears a deeper physical meaning which should manifest independently of the adopted measure.
Answering this question is the purpose of our work.

We consider a selection of essentially all the known {\it bona fide} measures of general quantum correlations defined in recent literature, and we find that they {\it all} freeze under the same dynamical conditions. A geometric analysis, inspired by \cite{caves}, is carried out in order to provide a satisfactory interpretation to the universality of this dynamical phenomenon. Our conclusions demonstrate that for all quantum information protocols (e.g.~remote state preparation, entanglement distribution, or quantum correlations-assisted parameter estimation \cite{datta,madhok:pr2011a,cavalcanti:pr2011a,np1,modix,lqu,distrib1,distrib2,distrib3}) whose performance relies on some form of discord between two qubits, however quantified, there exist unique noisy evolutions in the state space for which coherence and thus quantumness in the correlations are exactly preserved. The corresponding protocols will then run with efficiency unaffected by such noisy conditions.

\bigskip

The paper is organized as follows. In Section~\ref{secBell} we briefly recall the description of Bell diagonal states of two qubits. In Section~\ref{secDM} we present a comprehensive compendium of the measures of nonclassical correlations studied in this work.  Section~\ref{secU} demonstrates that all the considered measures froze under specifical nondissipative dynamical trajectories. In Section~\ref{secG} we present a geometric interpretation of the phenomenon. Discussions and conclusion are provided in Section~\ref{secC}, while some technical proofs are deferred to Appendices.

\section{Bell Diagonal states}\label{secBell}
We focus our attention to Bell diagonal (BD) states of two qubits, i.e., states with maximally mixed marginals \cite{horodecki1996information,entanglement}. Their density matrix can be written in Bloch form as $\rhobd=\frac 14(\id_{AB}+\sum_{i=1}^3 c_i \sigma_i^A\otimes \sigma_i^B)$ where $\{\sigma_i^{A,B}\}$ denote the Pauli matrices and the vector $\vec{c} = (c_1,c_2,c_3)$ completely specifies the state.

Explicitly,
\begin{equation}\label{rho}
\rhobd=\frac14
\left(
\begin{array}{cccc}
 c_3+1 & 0 & 0 & c_1-c_2 \\
 0 & 1-c_3 & c_1+c_2 & 0 \\
 0 & c_1+c_2 & 1-c_3 & 0 \\
 c_1-c_2 & 0 & 0 & c_3+1 \\
\end{array}
\right).
\end{equation}
As the name suggests, BD states $\rhobd$ have the four maximally entangled Bell states as eigenvectors, with eigenvalues \begin{equation}\label{automoto}
\lambda^{\vec{c}}_{ab}=\mbox{$\frac14$} \big[1+(-1)^a c_1 -(-1)^{a+b} c_2 + (-1)^b c_3\big]\,,
\end{equation}
where $a,b=0,1$. The conditions $\lambda^{\vec{c}}_{ab} \geq 0$ impose constraints on the entries of $\vec{c}$, so that physically allowed BD states can be represented as points within a tetrahedron of vertices $(-1,-1,-1)$, $(-1,1,1)$, $(1,-1,1)$ and $(1,1,-1)$ in the three-dimensional space spanned by $(c_1,c_2,c_3)$ \cite{horodecki1996information}. This geometric picture is very appealing to visualize dynamical trajectories \cite{caves}.
For later convenience, we define an auxiliary ordered vector $\vec{\varsigma}=(\varsigma_1 \equiv |c_{l_1}|,\varsigma_2 \equiv |c_{l_2}|, \varsigma_3 \equiv |c_{l_3}|)$, where $\{l_1,l_2,l_3\}$ is a suitable permutation of $\{1,2,3\}$ such that $\varsigma_1 \geq \varsigma_2 \geq \varsigma_3$. BD states with zero discord correspond then to $\varsigma_2=\varsigma_3=0$.

\section{Discord measures}\label{secDM}
We study one-way measures $\cal Q$ of discord, that is, those which reveal quantumness of correlations as perceived by an observer probing only qubit $A$. Incidentally, whenever symmetrized versions of these measures are available \cite{modirev}, they coincide with their one-way counterparts for BD states, thus enlarging the scope of our study. The adopted measures, which include those reviewed in \cite{modirev}, others taken from recent literature, and some defined or calculated here for the first time, are listed in the following \cite{notenorm}. Most of these measures do not enjoy a closed analytical form in general, but are computable for BD states. Notice that pairs of distinct quantifiers selected from our list will generically induce inequivalent orderings on the set of non-CQ states, even within the BD class \cite{malgorzata}.

\subsection{Quantum discord ${\cal D}$}
The original measure of quantum discord \cite{OZ,HV} can be defined as the minimum difference in total correlations between the state $\rho$ and the CQ state obtained after an optimized projective measurement $\{\Pi^A_i\}$ on $A$, ${\cal D}(\rho_{AB})=\min_{\{\Pi^A_i\}}\big[{\cal I}(\rho_{AB})-{\cal I}(\sum_i \Pi^A_i {\rho}_{AB} \Pi^A_i )\big]$. Here ${\cal I}(\rho_{AB})=S(\rho_A)+S(\rho_B)-S(\rho_{AB})$ is the mutual information \cite{groisman2005quantum}, with $S(\rho)=-\text{Tr}(\rho \log_2 \rho)$ being the von Neumann entropy. For BD states, $S(\rho^{\vec{c}}_A)=S(\rho^{\vec{c}}_B)=1$ while $S(\rhobd)=-\sum_{a,b=0}^1 \lambda_{ab}^{\vec{c}} \log_2 (\lambda_{ab}^{\vec{c}})$. The discord can be computed as
\cite{luobd,caves}
 ${\cal D}(\rhobd) = 1 -S(\rhobd)+ H\left(\frac{1+\varsigma_1}{2}\right)$, where $H(s)=-s \log_2 s - (1-s) \log_2(1-s)$.

\subsection{One-way quantum deficit $\Delta^{\rightarrow}$}
This measure  \cite{deficit1,deficit2} quantifies, in a thermodynamical framework, the minimum entropy production after a projective measurement on $A$, $\Delta^{\rightarrow}(\rho_{AB}) = \min_{\{\Pi^A_i\}}\big[S(\sum_i \Pi^A_i {\rho}_{AB} \Pi^A_i)-S(\rho_{AB})\big]$. It can be interpreted as the amount of information in the state $\rho_{AB}$ which cannot be localized via a one-way channel of classical communication from $A$ to $B$. For BD states, $\Delta^{\rightarrow}(\rhobd)={\cal D}(\rhobd)$ \cite{luobd}.

\subsection{Relative entropy of discord $D_R$}
This quantifier  \cite{deficit2,groisreq,modi} captures  the distance, as measured by the quantum relative entropy \cite{vedralrev}, between $\rho_{AB}$ and the set of CQ states, $D_R(\rho_{AB})=\min_{\chi_{AB} \in \ohm} S(\rho_{AB}||\chi_{AB})$, where $S(\rho||\sigma)=\text{Tr}[\rho(\log_2 \rho - \log_2 \sigma)]$. For all bipartite states, $D_R(\rho_{AB})=\Delta^{\rightarrow}(\rho_{AB})$ \cite{deficit2,modi}.

\subsection{Adjusted geometric discord $\widetilde D_G$}
The geometric discord $D_G$ was defined in \cite{dakic,luofu} as the minimum squared Hilbert-Schmidt distance from the set $\Omega_{\rm CQ}$. However, it is now acknowledged that such a definition is flawed \cite{piani}, as $D_G$ can be modified arbitrarily by reversible operations on $B$. Moreover, $D_G$ does not respect {\it P3} \cite{cinesi,tufoibrido}. It is customary, when dealing with the Hilbert-Schmidt distance, to normalize it by the purity of the state, in order to correct the metric for the effective dimension of the Hilbert space \cite{winter,genoni}. We then define the adjusted geometric discord as $\widetilde{D}_G(\rho_{AB}) = 2 \big(\min_{\chi_{AB} \in \ohm} \|\rho_{AB} - \chi_{AB}\|_2^2\big)/\text{Tr}(\rho_{AB}^2)$, where $\|M\|^2_2=\text{Tr}[M^\dagger M]$. Properties and applications of this measure are discussed in  \cite{tuford}. For BD states, using \cite{dakic} we have $\widetilde{D}_G(\rho^{\vec{c}}_{AB})=2(\varsigma_2^2+\varsigma_3^2)/(1+\varsigma_1^2+\varsigma_2^2+\varsigma_3^2)$.

\begin{figure*}[t]
\hspace*{-.1cm}
\subfigure[]{
\includegraphics[width=3.4cm]{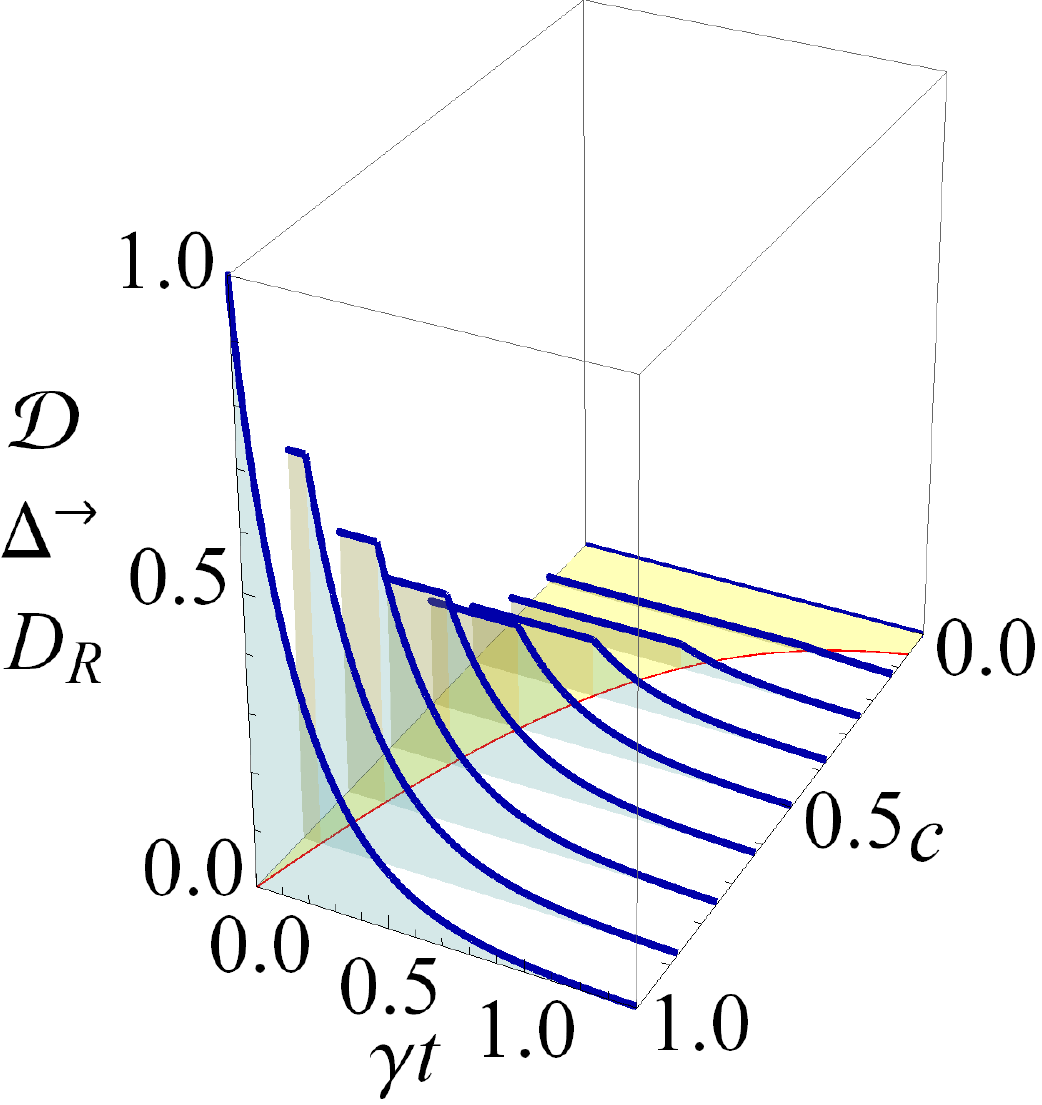}\label{plD}}\hspace*{.1cm}
\subfigure[]{
\includegraphics[width=3.4cm]{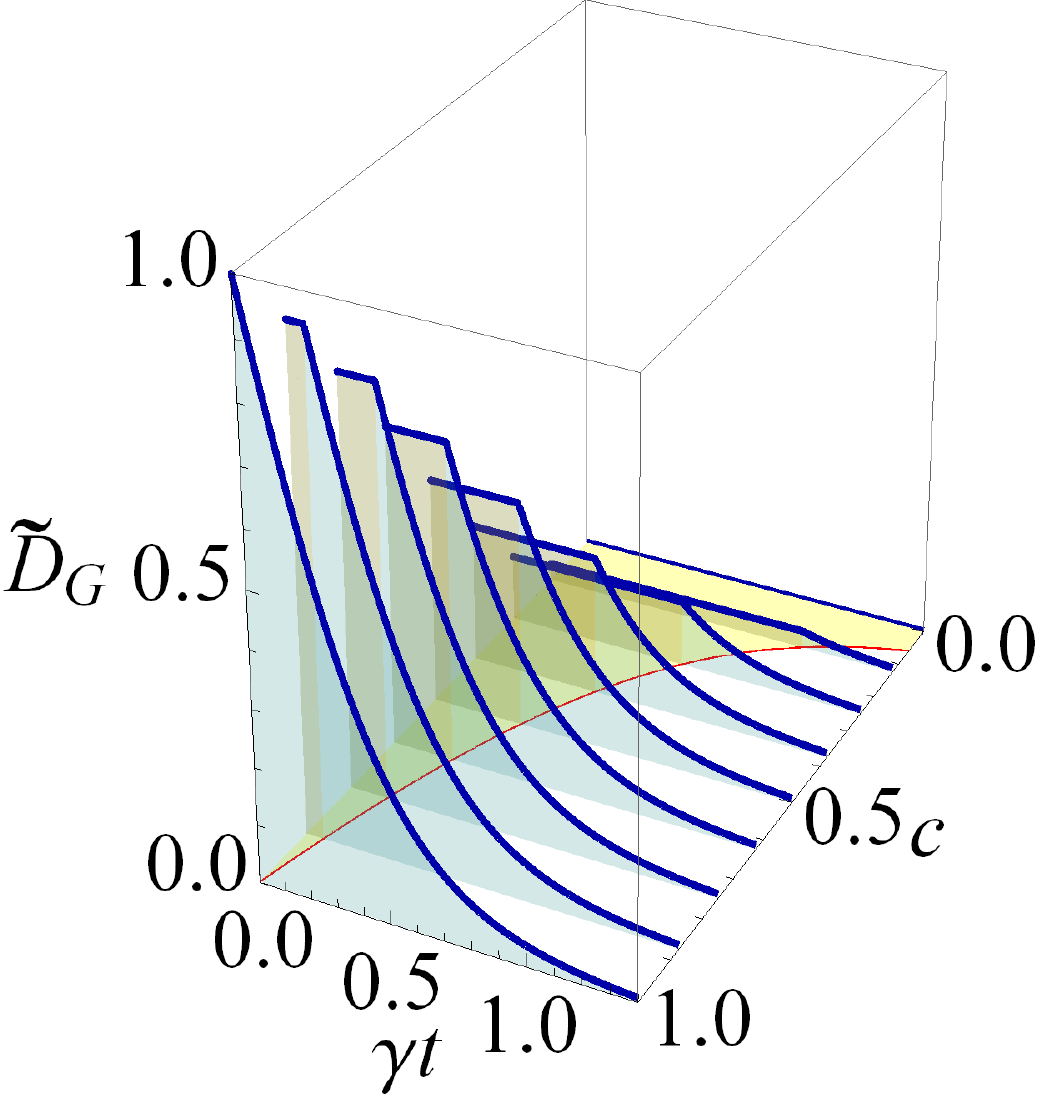}\label{plG}}\hspace*{.1cm}
\subfigure[]{
\includegraphics[width=3.4cm]{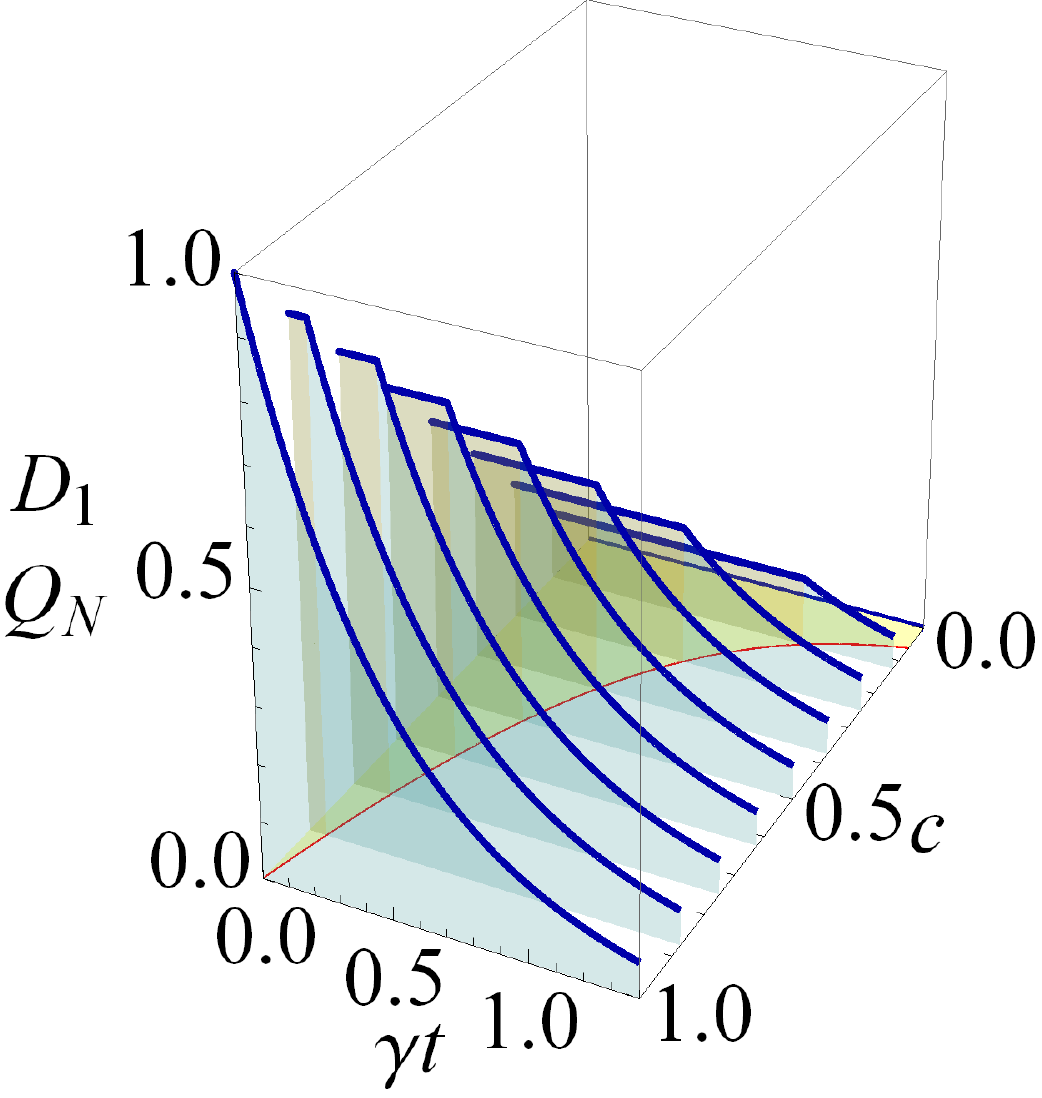}\label{plN}}\hspace*{.1cm}
\subfigure[]{
\includegraphics[width=3.4cm]{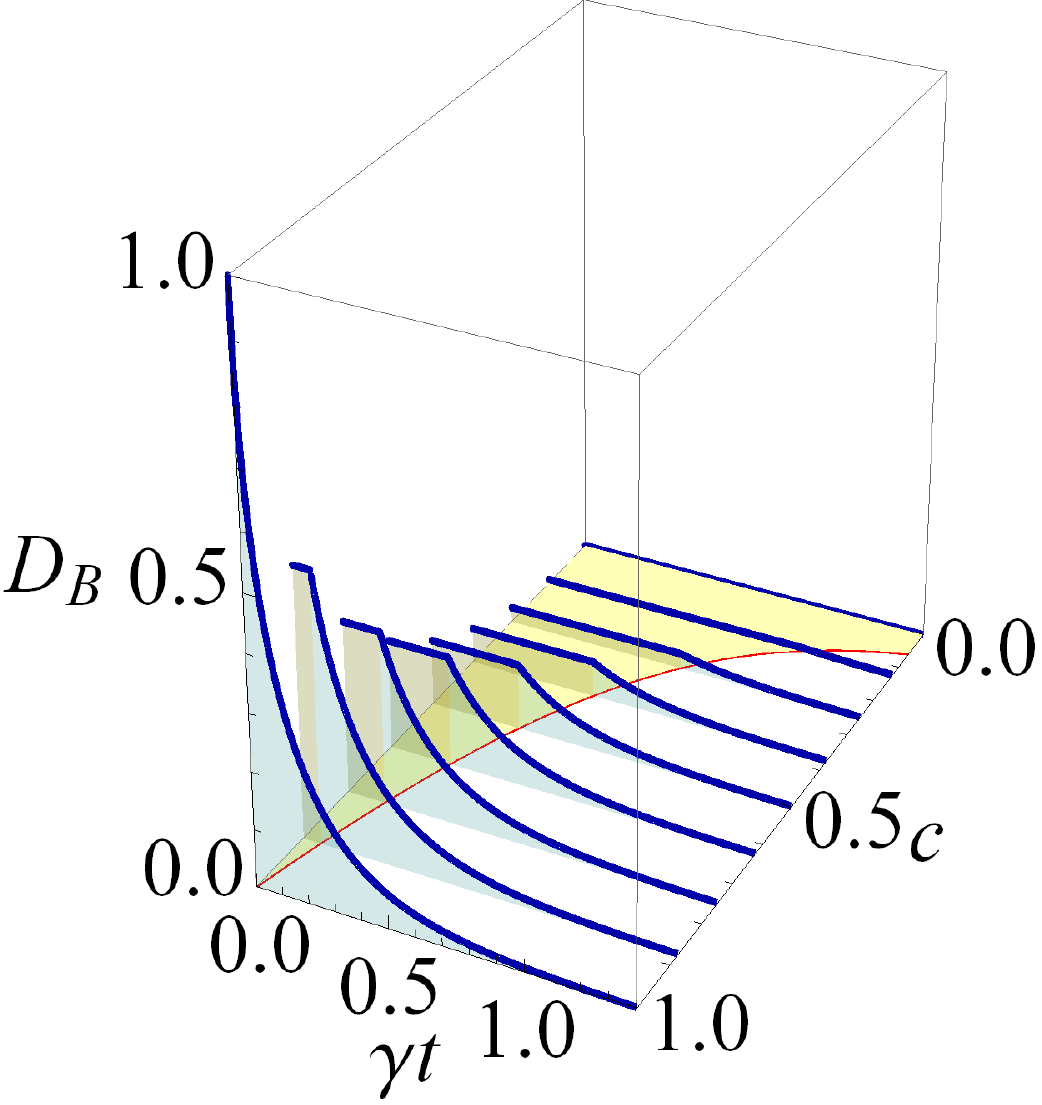}\label{plB}}\hspace*{.1cm}
\subfigure[]{
\includegraphics[width=3.4cm]{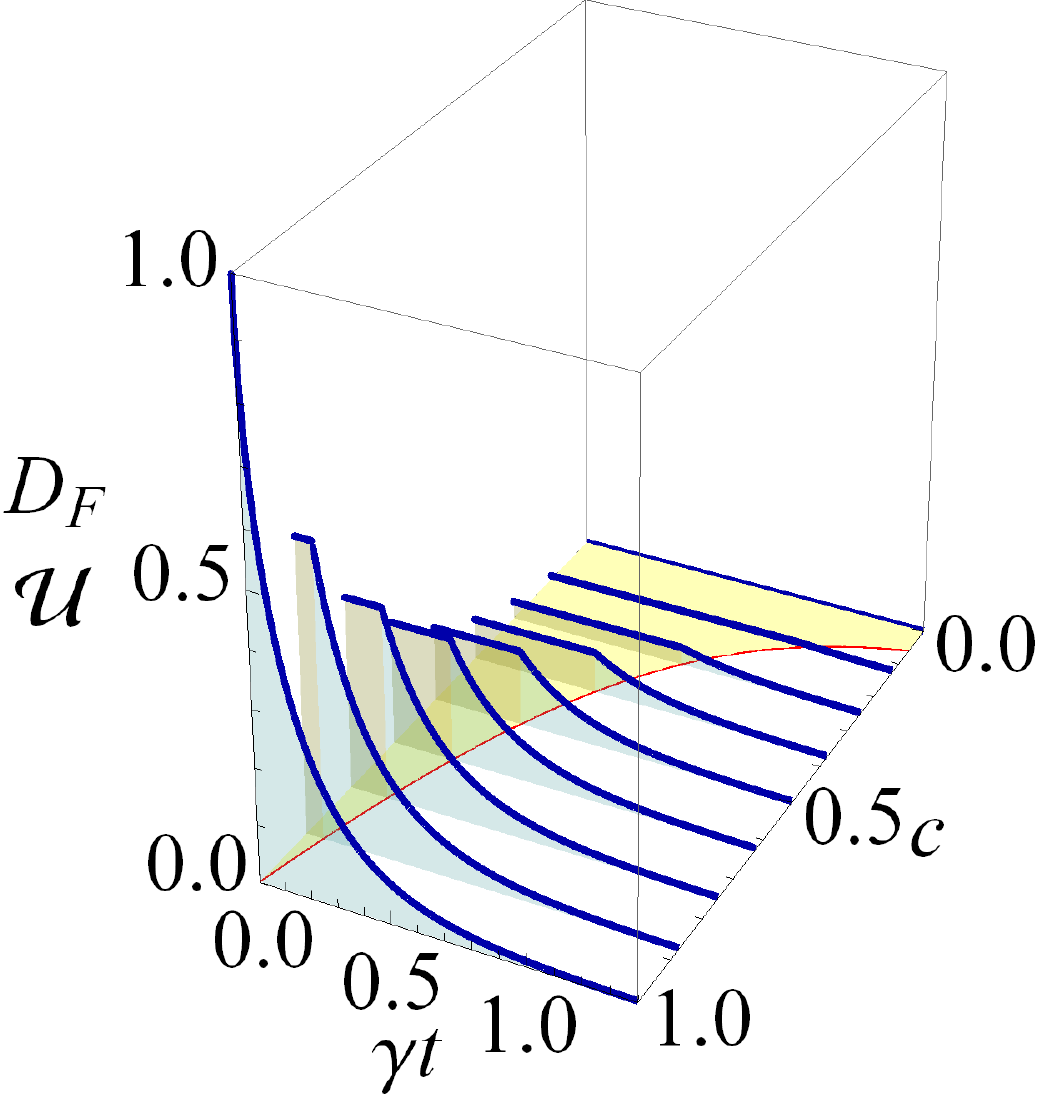}\label{plU}}
\caption{(Color online). Time evolution of various measures of quantum correlations for BD states with initial condition $\vec{c}^{(k)}(0)$ given by Eq.~(\ref{rihanna}), subject to local bit flip ($k=1$), bit-phase flip ($k=2$) or phase flip ($k=3$) channels with rate $\gamma$. In the range of parameters such that $2\gamma t < - \ln c$, depicted as a shaded (yellow online) region on the horizontal planes in each panel, all the considered measures remain frozen to their respective initial values. The plotted quantities are: (a) quantum discord ${\cal D}$, equal to one-way quantum deficit $\Delta^{\rightarrow}$, equal to relative entropy of discord $D_R$; (b) adjusted geometric discord $\widetilde{D}_G$; (c) trace-distance discord $D_1$, equal to negativity of quantumness $Q_N$; (d) Bures-distance discord $D_B$; (e) fidelity-based measure $D_F$, equal to local quantum uncertainty ${\cal U}$. The analytical expressions are given by combining Eq.~(\ref{allfrozen}) with Table~\ref{tablef}.}
\label{figafresca}
\end{figure*}

\subsection{Trace-distance discord $D_1$}
As the name suggests, this measure \cite{taka,vianna,newonenorm} is given by the minimum trace distance from the set of CQ states, $D_1(\rho_{AB})=\min_{\chi_{AB} \in \ohm} \|\rho_{AB} - \chi_{AB}\|_1$, with $\|M\|_1=\text{Tr}\big[\sqrt{M^\dagger M}\big]$. For BD states, such a measure is computable \cite{taka,newonenorm,giovannetti} and one has simply $D_1(\rhobd)=\varsigma_2$.

\subsection{Negativity of quantumness $Q_N$}
This measure  \cite{activation,taka} corresponds to the minimum negativity (an entanglement measure \cite{negativity}) created between the system $AB$ and an apparatus $C$ during a local projective measurement on $A$, according to the formalism of \cite{streltsov,activation,pianiadesso}. Referring the reader to \cite{taka,newonenorm} for details, we recall that if subsystem $A$ is a qubit (as in our case), then $Q_N(\rho_{AB}) = D_1(\rho_{AB})$.

\subsection{Bures-distance discord $D_B$}
We can consider the Bures distance from the set of CQ states as another measure of discord, in analogy with the Bures measure of entanglement \cite{buresent}. We  then have $D_B(\rho_{AB})=
\big[\big(2+\sqrt{2}\big) \big(1-\sqrt{F_{\max}(\rho_{AB})}\big)\big]^{\frac12}$, where $F_{\max}(\rho_{AB})=\max_{\chi_{AB} \in \ohm} F(\rho_{AB},\chi_{AB})$ and  $F(\rho,\sigma)=\big\{\text{Tr}\big[\big(\sqrt{\rho} \sigma \sqrt{\rho}\big)^{\frac12}\big]\big\}^2$ is the Uhlmann fidelity \cite{fidelity}. For BD states, in Appendix A we obtain an analytical expression for $F_{\max}$  (see \cite{fabernew} for an independent yet related derivation). We notice that, for a given $\rhobd$, the CQ state which maximizes the fidelity is in general different from the one which minimizes trace distance, Hilbert-Schmidt distance, and relative entropy. Explicitly, we have $F_{\max}(\rhobd)=\frac12+\frac14\max_{\langle i,j,k\rangle} \big[
\sqrt{(1+c_i)^2-(c_j-c_k)^2}+\sqrt{(1-c_i)^2-(c_j+c_k)^2}\big]$, where $\langle i,j,k\rangle$ denotes cyclic permutations of $\{1,2,3\}$.

\subsection{Fidelity-based measure $D_F$}
As a simple rescaling of the previous quantity, we also pick the geometric measure of quantumness discussed in \cite{streltsov,behavior,tahere}, which in our notation reads $D_F(\rho_{AB})=2[1-F_{\max}(\rho_{AB})]$, and is computable for BD states using the expression for $F_{\max}$ given above.

\subsection{Local quantum uncertainty ${\cal U}$}
The last measure we adopt, which is also the most recently introduced  \cite{lqu}, quantifies the minimum quantum uncertainty on a single local observable, ${\cal U}(\rho_{AB}) = \min_{K_A} I_{\rm WY}(\rho_{AB}, K_A \otimes \id_B)$, where $K_A$ is a Hermitian operator with nondegenerate spectrum acting on subsystem $A$, and $I_{\rm WY}(\rho,K)=\text{Tr}\big[\rho K^2 - \sqrt{\rho} K \sqrt{\rho} K\big]$ is the Wigner-Yanase skew information \cite{wigner}. The quantity ${\cal U}$ is computable for all two-qubit states
\cite{lqu}. Interestingly, we find that for BD states ${\cal U}(\rhobd) = D_F(\rhobd)$, although the two measures do not coincide in general.

\section{Universal freezing}\label{secU}
We consider local Markovian nondissipative decoherence channels acting independently on each of two qubits $m=A,B$ in the state $\rho_{AB}$. The single-qubit Lindblad operator has the form
\begin{equation}{\cal L}_k[\rho_m]=\frac{\gamma}{2}
\big(\sigma_k^m \rho_m \sigma_k^m - \rho_m\big),\end{equation}
where $\gamma$ is the decoherence rate and $k=1,2,3$ denote bit flip, bit-phase flip, and phase flip (alias phase damping) channels, respectively \cite{maziero,mazzola}.
We consider BD states $\rho_{AB}^{\vec{c}(0)}$ as inputs.
The time-evolved states under the considered local channels remain in BD form, with
 \begin{equation}\label{evolvc}
c_{i,j \neq k}(t)=c_{i,j}(0) e^{-2\gamma t},\,c_k(t)=c_k(0)\,,
\end{equation}
where $k$ selects the channel as explained above.
Different initial conditions $\vec{c}(0)$ lead to varied dynamics of nonclassical correlations, without any general agreement between the measures we consider. However, for each selected channel, indexed by $k$, we can choose a specific subset of initial conditions $\vec{c}^{(k)}(0)$ \cite{mazzola}, given by
\begin{equation}\label{rihanna}
\mbox{$c^{(k)}_i(0)=\pm 1,\,c^{(k)}_j(0) = \mp c^{(k)}_k(0)$, with $|c^{(k)}_k(0)|\equiv c$}
 \end{equation}
 (notice that taking into account the sign freedom and the permutation of $i,j\neq k$ there are four possible choices per channel). These conditions, which are equivalent to imposing that the density matrix be of rank $2$, give rise to very peculiar dynamics of the quantum correlations in the time-evolved states $\rho_{AB}^{\vec{c}^{(k)}(t)}$ under the corresponding $k$-type channels. Namely, defining the threshold time $\gamma t^{\star}=-\frac12  \ln c$, we find analytically that every measure ${\cal Q}$ considered in this paper takes the form
\begin{equation}\label{allfrozen}
{\cal Q}\big(\rho_{AB}^{\vec{c}^{(k)}(t)}\big)=\left\{
                   \begin{array}{ll}
                     f_{\cal Q}(c), & \hbox{if $0\leq t < t^{\star}$;} \\
                     f_{\cal Q}\big(e^{-2 \gamma t}\big), & \hbox{if $t \geq t^{\star}$.}
                   \end{array}
                 \right.
\end{equation}
This entails that {\it all} studied measures of discord remain {\it frozen} to their initial value until, at $t=t^{\star}$,  they suddenly start decaying exponentially, as depicted in Fig.~\ref{figafresca}. The  functional dependence $f_{\cal Q}$ for each measure ${\cal Q}$ is reported in Table~\ref{tablef}.

\begin{table}[t]
\begin{tabular}{cc}
  \hline \hline
Measure  ${\cal Q}$ & $f_{\cal Q}(s)$  \\ \hline
  Quantum discord ${\cal D}$ & $\frac12\!\sum_{a=0}^1[1+(-1)^a s] \log_2[1+(-1)^a s]$ \\
  One-way quantum deficit $\Delta^{\rightarrow}$ & $\frac12\!\sum_{a=0}^1[1+(-1)^a s] \log_2[1+(-1)^a s]$ \\
  Relative entropy of discord $D_R$ & $\frac12\!\sum_{a=0}^1[1+(-1)^a s] \log_2[1+(-1)^a s]$ \\
  Adjusted geometric discord  $\widetilde{D}_G$ & $2s^2/(1+s^2)$ \\
  Trace-distance discord $D_1$ & $s$ \\
  Negativity of quantumness $Q_{N}$ & $s$ \\
  Bures-distance discord $D_B$ & $1+\big(1+\sqrt{2}\big) \big[1-\big({\sqrt{1-s^2}+1}\big)^\frac12\big]$ \\
  Fidelity-based measure $D_F$ & $1-\sqrt{1-s^2}$ \\
  Local quantum uncertainty ${\cal U}$ & $1-\sqrt{1-s^2}$ \\
  \hline \hline
\end{tabular}
\caption{Summary of the expressions for the measures ${\cal Q}$ of quantum correlations in the BD states with initial condition $\vec{c}^{(k)}(0)$ evolving under local nondissipative channels, as evaluated in Eq.~(\ref{allfrozen}). \label{tablef}}
\end{table}

The presented results, although tied to the particular choices of channels and initial states, apply to  such a variety of measures of discord that the level of mere coincidence can be safely considered overcome. We now seek for a physical explanation of our findings. Most of our employed measures, in the chosen settings, can be interpreted as minimal distances from the set of CQ states. It turns out that, for nondissipative evolutions with initial condition as in Eq.~(\ref{rihanna}), all our distance functions are optimized by the same time-dependent CQ state (explicitly given in \cite{mazzola,lauraint,sarogeom}). This certainly explains why all such measures have a sudden change at the same time $t=t^{\star}$, but does not explain why all of them are in fact {\it constant} for earlier times. For instance, without the purity adjustment, the geometric discord $D_G$ \cite{dakic} decreases rather than being frozen when $t<t^{\star}$ \cite{xu,sarogeom,cavesgeom}. The adjusted $\widetilde{D}_G$ instead freezes like the other measures. As remarked earlier, the purity adjustment is necessary to correct for some flaws of $D_G$ \cite{piani,tuford}. This suggests that the occurrence of freezing is, at a somewhat empirical level, a stronger feature which necessitates to be present in truly {\it bona fide} quantifiers of discord.

\begin{figure}[tb]
\includegraphics[width=8.5cm]{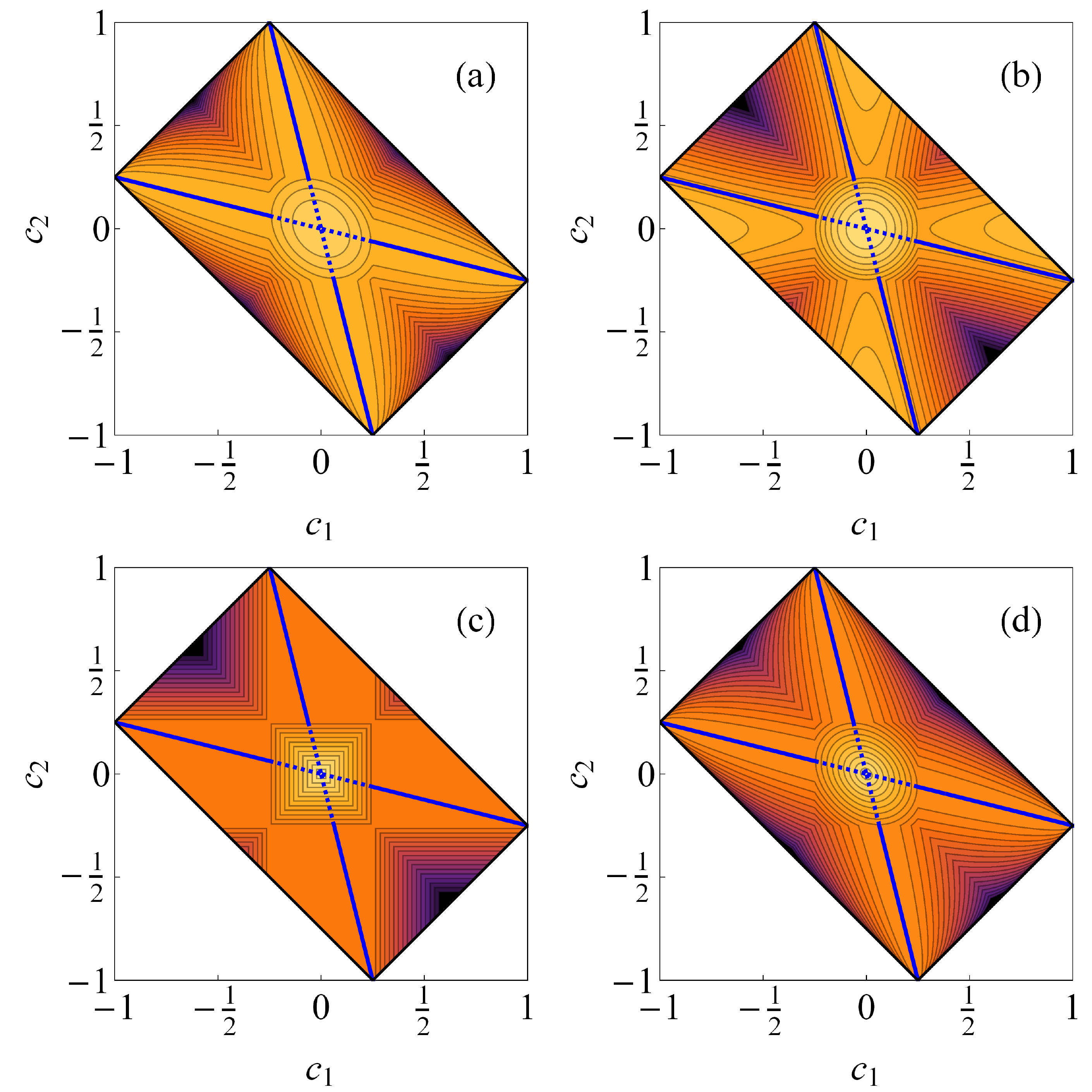}
\caption{(Color online). Contours of constant ${\cal Q}$ in the plane $(c_1,c_2)$ for physical BD states with $c_3=\frac14$, where ${\cal Q}$ represents (a) ${\cal D}\equiv \Delta^{\rightarrow} \equiv D_R$, (b) $\widetilde{D}_G$, (c) $D_1\equiv Q_N$, (d) $D_B$. The diagonals (traced from the vertices to the centre) represent the evolution of BD states with initial condition $\vec{c}^{(3)}(0)$ [Eq.~\ref{rihanna}] under local phase flip channels. For $t< t^{\star}$ (continuous traits) all the measures ${\cal Q}$ are constant; they then decay to zero for later times (dashed traits). The color legend for the ${\cal Q}$'s in the contour plots is: $0$ \protect\includegraphics[width=1.3cm]{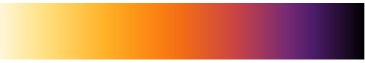} $1$. }
\label{sconcia}

\end{figure}

\section{Geometric representation}\label{secG}
A deeper insight into this phenomenon can then be achieved by looking at the geometric representation of BD states, similarly to what was done in \cite{caves} and in \cite{cavesgeom} for the quantum discord ${\cal D}$ and for the (non-adjusted) geometric discord $D_G$, respectively. In the space $(c_1,c_2,c_3)$, one can draw the surfaces along which each given measure ${\cal Q}$ is constant, and then superimpose the trajectories corresponding, say, to the evolution of $\vec{c}^{(k)}(t)$ under local $k$-channels, in order to visualize the freezing. Let us focus, without loss of generality, on phase flip channels $(k=3)$.
For ease of graphical display, instead of drawing three-dimensional surfaces, we find it more informative to slice the tetrahedron of BD states at, say, a constant value of $c_3$, and draw the contours of constant ${\cal Q}$ in the resulting projected plane spanned by $(c_1,c_2)$. This analysis is reported in Fig.~\ref{sconcia} for several types of discord measures. It is evident that the various quantifiers exhibit their quite different nature even within the restricted set of BD states. In particular, around the classical states at the centre $(c_1,c_2)=(0,0)$, contours of constant quantum correlations are ellipses for entropic (${\cal D}, \Delta^{\rightarrow}, D_R$) and fidelity-based ($D_B, D_F, {\cal U}$) measures, while they are circles for the Hilbert-Schmidt based $\widetilde{D}_G$, and squares for the trace-distance based $D_1$. Similarly, closer to the peripheral boundaries of the physically allowed region, the contour lines have different topologies across the various measures. Some measures like $D_1$ are constant for wider regions than the others, for instance.
Here we are interested in the overlap between the contour lines of all ${\cal Q}$'s.
We find (see Appendix B) that the only possible straight lines, in the BD state space, which keep {\it all} the considered measures simultaneously constant, are those along the diagonals of the projected rectangles in the space $(c_1,c_2)$, stopping sufficiently far away from the central core (see Fig.~\ref{sconcia}). These segments describe only and precisely the evolutions of BD states with initial conditions $\vec{c}^{(3)}(0)$ [Eq.~(\ref{rihanna})] for local phase flip channels---and similarly for the other $k$-channels by selecting the corresponding planes $(c_i,c_j)$ and conditions $\vec{c}^{(k)}(0)$---up to the time $t^{\star}$. This yields a general {\it geometric interpretation} to the phenomenon of frozen discord universally observed in all valid measures at once.

\section{Discussion and concluding remarks}\label{secC}
We wish to remark that, in the presence of suitable non-Markovian channels which can be described by a master equation with a memory kernel (as in the case of pure dephasing or decoherence under classical random external fields) \cite{lauraint,pigna,sarogeom,sarophys,sarorevival}, all the phenomena previously observed for the entropic discord ${\cal D}$ extend automatically to all the measures considered in this paper. Indeed, in the models of \cite{lauraint,pigna,sarogeom,sarophys,sarorevival}, the evolution of BD states can be formally written as in Eq.~(\ref{evolvc}), but with $2 \gamma t$ replaced by a more general $\Lambda(t)$, which could be decreasing over some time intervals. This can give rise to dynamics with multiple intervals of constant discord \cite{lauraint,sarogeom,sarorevival}, or discord frozen forever \cite{pigna} depending on the initial conditions. By our analysis, we conclude that those features, which might be observable e.g.~in the dynamics of impurity atoms in Bose-Einstein condensates \cite{pigna,gabriele}, are universal too and detectable by any suitable discord measure ${\cal Q}$.

In this paper we established the general status of an intriguing aspect of quantum correlations other than entanglement: the {\it freezing} in certain dynamical conditions. Originally revealed for entropic quantifiers \cite{mazzola}, we showed that this feature is common to all {\it bona fide} measures of discord, and we provided a geometric interpretation thereof.
It will be interesting to investigate, when the theory of discord \cite{modirev} is completed, whether the occurrence of freezing in nondissipative evolutions will perhaps be provable  as an implication of the  set of necessary conditions for assessing the mathematical validity of measures of discord. At present, the occurrence of freezing in specific nondissipative evolutions can be proposed as a `sanity check' to validate novel discord quantifiers. From an operative perspective, we showed that quantumness of correlations, in all its manifold  manifestations, can be sustained with no loss in suitable noisy settings. Our predictions are amenable to experimental verification  with current technology, e.g.~using photons \cite{guo} or nuclear magnetic resonance techniques  \cite{braziprl,braziprlours}. We expect our study to stimulate novel endeavors in the comprehension and exploitation of genuinely quantum effects in open systems.

\section*{ACKNOWLEDGEMENTS}
We acknowledge fruitful discussions with F. Ciccarello, G. Compagno, L. Correa, D. Girolami, P. Haikka, F. Illuminati D. Soares-Pinto, A. Streltsov, T. Tufarelli. We acknowledge financial support from the University of Nottingham through an Early Career Research and Knowledge Transfer Award and an EPSRC Research Development Fund Grant (PP-0313/36).

\appendix

\section{Fidelity-based measure of discord}
A two-qubit state $\rho$ can be written in the Bloch representation:
\begin{equation}
\rho = \frac{1}{4}(\id^A \otimes \id^B + \sum\limits_{i=1}^3 x_i \sigma_i^A \otimes \id^B + \sum\limits_{i=1}^3 y_i \id^A \otimes \sigma_i^B + \sum \limits_{i,j=1}^3 T_{ij} \sigma_i^A \otimes \sigma_j^B)
\end{equation}
where $\{\sigma_i^{A,B}\}$ denote the Pauli matrices. With such a state $\rho$ we associate the triple  $\{\vec{x}, \vec{y},\textbf{T}\}$  \cite{dakic}.
\\
\\
{\bf Theorem 1.} {\it For any Bell diagonal (BD) state $\rho$, $F(\rho, \sigma_0) \geq F(\rho, \sigma)$, where $\sigma$ is the state with triple  $\{\vec{x}, \vec{y},\textbf{T}\}$, and  $\sigma_0$ the state with triple $\{\vec{0}, \vec{0},\textbf{T}\}.$}

{\it Proof.} For any state $\sigma$, there exists another state with associated triple $\{-\vec{x}, -\vec{y},\textbf{T}\}$ which we denote $\sigma_-$. It can be verified that for BD $\rho$, the matrices $\sqrt{\rho}\sigma\sqrt{\rho}$ and $\sqrt{\rho}\sigma_-\sqrt{\rho}$ have the same characteristic polynomial, and so have the same eigenvalues. From the expression for fidelity
\begin{equation}
F(\rho,\sigma) = \tr{\sqrt{\sqrt{\rho}\sigma\sqrt{\rho}}}
\end{equation}
it can be seen that the equality of the eigenvalues is sufficient to prove the identity
\begin{equation} \label{plusminusequal}
F(\rho,\sigma) = F(\rho,\sigma_-).
\end{equation}

Fidelity obeys the concavity property
\begin{equation}
F(\rho, \mu \sigma_1 + (1-\mu) \sigma_2) \geq \mu F(\rho, \sigma_1) + (1-\mu) F(\rho, \sigma_2), \mu \in [0,1]
\end{equation}
Setting $\mu = \frac{1}{2}$, $\sigma_1 = \sigma$, $\sigma_2 = \sigma_-$,  using Eq.~(\ref{plusminusequal}) and noting that $\frac{1}{2}(\sigma + \sigma_-) = \sigma_0$, we recover $F(\rho, \sigma_0) \geq F(\rho, \sigma)$ and so prove the theorem.  \hfill $\Box$

Equipped with this, it is now possible to prove the main result.
\\
\\
{\bf Theorem 2.} {\it For any BD state $\rho$, the classical-quantum (CQ) state $\chi$ which maximizes $F(\rho,\chi)$ is also BD}

{\it Proof.}
Any CQ two-qubit state will be of the form $\chi = p \ketbra{\psi_1}^A \otimes \rho_1^B + (1-p) \ketbra{\psi_2}^A \otimes \rho_2^B$, where $\{\ket{\psi_1}^A,\ket{\psi_2}^A\}$ is an orthonormal basis for qubit $A$. Such a CQ state will have associated triple $\{(2p-1)\vec{e}, \vec{s}_+, \vec{e}\vec{s}_-^T\}$ \cite{dakic}, where

\begin{eqnarray}
e_i &=& \bra{\psi_1}\sigma_i \ket{\psi_1},\\
s_{\pm,i} &=& \tr{(p \rho_1 \pm (1-p)\rho_2)\sigma_i}.
\end{eqnarray}
For any state in this form, a second state $\chi_0 =  p' \ketbra{\psi_1}'^A \otimes \rho_1'^B + (1-p') \ketbra{\psi_2}'^A \otimes \rho_2'^B$ can be derived using the identity
\begin{subequations}
\begin{eqnarray}
p' = \frac{1}{2} \\
\ket{\psi_1}'^A = \ket{\psi_1}^A \\
\ket{\psi_2}'^A = \ket{\psi_2}^A \\
\rho_1'^B = \frac{1}{2}(\id^B + p \tau_1^B - (1-p)\tau_2^B) \\
\rho_2'^B = \frac{1}{2}(\id^B -p \tau_1^B + (1-p)\tau_2^B)
\end{eqnarray}
\end{subequations}
where $\tau_1, \tau_2$ are the traceless parts of $\rho_1, \rho_2$. This state is manifestly CQ, and it can be easily verified that it will have associated triple $\{\vec{0}, \vec{0},\textbf{T}\}$. From Theorem 1. we can see that it suffices to restrict ourselves to CQ states of this form.

Temporarily, we relax the restriction that $\vec{e}$ is of unit length, and consider the set of states where $\|\vec{e}\| \leq 1$.  This allows us to repeat the previous trick, this time between $\chi$ with $\vec{e} = (e_1,e_2,e_3), \vec{s_-} = (s_1, s_2, 0)$ and $\chi'$ with $\vec{e'} = (e_1,e_2,-e_3), \vec{s'_-} = \vec{s_-}$. As before, from the comparison of characteristic polynomials, $F(\rho, \chi) = F(\rho, \chi')$. A similar result holds for $s_1 = 0$ and $s_2 = 0$ and, by switching the vectors we consider, for any $e_i=0$. From this we can see that if any $s_i = 0$ then for maximum fidelity, $e_i = 0$, and vice versa. We also note that this set of states is convex, and so due to the concavity of fidelity, any local maximum will be a global maximum.

As an ansatz, we now consider the states where $e_j = e \delta_{ij}, s_j = s \delta_{ij}$ where $i$ sets the non-zero vector element. From the previous result we can see that maximization only needs to be performed over $e_i$ and $s_i$ as fidelity can only decrease under any variation in any single other element. Furthermore, $e_i$ and $s_i$ appear only as a product $e_i s_i$ in the density matrix, never on their own. This means that maximizing over both is equivalent to setting $e_i=1$ and maximizing only over $s_i$, thus allowing us to reimpose the restriction that $\|\vec{e}\| = 1$ and returning to the CQ states. It is now only necessary to maximize over a single parameter, and any local maximum in this parameter will be the global maximum for CQ states. The remaining states are the BD CQ states, and so finding the maximum among these states proves the theorem.\\

Maximizing over the single remaining parameter we obtain the result for BD states with eigenvalues $\alpha, \beta, \gamma, \delta$ [of the form as in Eq.~(\ref{automoto})]:
\begin{equation}
D_F(\rho) = min\{q_1, q_2, q_3\}
\end{equation}
where
\begin{subequations}
\begin{align}
q_1 = 1 - \sqrt{2}\left(\sqrt{\alpha \delta} + \sqrt{\beta \gamma}\right) \\
q_2 = 1 - \sqrt{2}\left(\sqrt{\alpha \gamma} + \sqrt{\beta \delta}\right) \\
q_3 = 1 - \sqrt{2}\left(\sqrt{\alpha \beta} + \sqrt{\gamma \delta}\right)
\end{align}
\end{subequations}

This is identical to the expression for local quantum uncertainty \cite{lqu} of BD states, giving the identity
\begin{equation}
D_F(\rho) = {\cal U(\rho)}
\end{equation}
for $\rho$ an arbitrary BD two-qubit state. \hfill $\Box$

\section{Universal freezing trajectories}
{\bf Theorem 3.} {\it Universal freezing for decay in the $k$-type flip channel is observed in BD states only for initial conditions of the form
\begin{equation} \label{startcond}
\mbox{$c^{(k)}_i(t_0)=\pm e^{-2\gamma t_0},\,c^{(k)}_j(t_0) = \mp c^{(k)}_k(t_0) e^{-2\gamma t}$, with $|c^{(k)}_k(t_0)|\equiv c$}
 \end{equation}}

{\it Proof.} We consider a BD state in the form $\rho_{AB}=\frac 14(\id_{AB}+\sum_{i=1}^3 c_i \sigma_i^A\otimes \sigma_i^B)$. For {\it universal} freezing, all {\it bona fide} measures must have a constant value up until the time $t^{\star}$. In particular, any two of them must satisfy this requirement. In the following, we will pick the negativity of quantumness and the adjusted geometric discord and find that the only straight-line freezing trajectory common to both is the one reported in the claim. Recall the expression for trace-distance discord ($D_1$) alias negativity of quantumness ($Q_N$), $D_1 = Q_N = \varsigma_2$, where $\{\varsigma_i\}$ are simply $\{|c_i|\}$ ordered such that $\varsigma_1 \geq \varsigma_2 \geq \varsigma_3$; from this it can immediately be seen all initial BD states $\rho$ satisfying
\begin{equation} \label{NoQCond}
\varsigma_2(t_0) = |c^{(k)}_k(t_0)| \equiv c
\end{equation}
will exhibit freezing for these measures.

The adjusted geometric discord $\widetilde{D}_{G}$ for the same states has the following expression for time $t \leq t^{\star}$:
\begin{equation}
\widetilde{D}_{G} = 2(c^{2} + \zeta_{3}^{2})/(1+ \zeta_{1}^{2} + c^{2} + \zeta_{3}^{2})
\end{equation}

Since $\varsigma_i(t) = \varsigma_i(t_0) e^{-2\gamma t}$ for $i =1, 3$, time evolution will follow a straight line $\ell$ in the $\zeta_{1} - \zeta_{3}$ plane. Such a line can be defined parametrically as:

\begin{subequations} \label{lineparam}
\begin{eqnarray}
\zeta_{1}(s) = s \\
\zeta_{3}(s) = ms+ a
\end{eqnarray}
\end{subequations}

In order for $\widetilde{D}_{G}$ to be constant over this time interval, $\ell$ must follow a contour line in this plane. Since contour lines are perpendicular to the gradient, it must therefore be:
\begin{equation}
\nabla \widetilde{D}_{G} . \nabla \ell = 0
\end{equation}
Calculating this and then substituting in the values for  $\zeta_{1}$ and $\zeta_{3}$ from Eq.~(\ref{lineparam}) gives:
\begin{equation}
a^2 s+a m \left(s^2-1\right)+s
  \left(c^2-m^2\right) = 0
\end{equation}
In order for $\ell$ to be a straight line, $m$ and $a$ must be independent of $s$. It can readily be verified by solving for $a$ that the only solution for which this is the case is $a=0, m = \pm c$.

Points on these lines correspond exactly to Eq.~(\ref{startcond}), proving the theorem. \hfill $\Box$


\begin{thebibliography}{72}
\bibitem[{not()}]{notenorm}
\bibinfo{note}{We include normalization factors in our definitions so that each
  measure reaches $1$ on maximally entangled Bell states.}
\end{thebibliography}
\end{document}